\documentclass[12pt,fullpage]{article}
\usepackage[english]{babel}
\usepackage{epsfig,epsf}
\usepackage{hyphenat}
\newcommand{\as}{\alpha_s}

\author{V.A. Abramovsky, N.V. Prikhod'ko \\
Novgorod State University, B. S.-Peterburgskaya Street 41,\\
Novgorod the Great, Russia, 117259}
\title{Energy dependence of Cronin momentum in saturation model for $p+A$ and $A+A$ collisions}

\begin{document}
\maketitle
\abstract{We calculate $\sqrt{s}$ dependence of Cronin momentum for $p+A$
and $A+A$ collisions in saturation model.
We show that this dependence is consistent with expectation from formula
which was obtained using simple dimentional consideration.
This can be used to test validity of saturation model
(and distinguish among its variants) and
measure $x$ dependence of saturation momentum from experimental data.
}
\section{Introduction}
It was shown in \cite{Baier:2003hr} that saturation model can explain Cronin like behavior.
As many other models also explain this behavior 
we need some subtle prediction to distinguish one model from another.
One prediction of this type is to calculate position of maximum in Cronin ratio.
But there are three variable in Cronin effect to measure: momentum where Cronin ratio have maximum 
(let's call it Cronin momentum $q_C$), value of maximum $R_C$ and momentum where Cronin ratio equal to unity $q_u$.
So why maximum position?
We know that value of Cronin ratio in $A+A$ collisions have normalization uncertainty.
And therefore variables $R_C$, $q_u$ are not a good ones to make predictions.
On the other side Cronin momentum does not depend on normalization and therefore is the best candidate we have.
Let's consider for now central rapidity $p+A$ collisions only. In saturation model there is only one semihard scale which govers
momentum dependence of differential cross-section $\frac{d\sigma_{pA}}{dydq^2}$ (here $y$ rapidity and $q$ transverce momentum of produced particles).
It is saturation momentum $Q_s$.
As we have only one semihard scale then Cronin momentum can only depend on
this scale. Using dimentional consideration the only equation which relates $Q_s$ and $q_C$ can be:
\begin{equation}
	q_C=\beta Q_s
	\label{pp_simple}
\end{equation}
where $\beta$ is some dimentionless constant.
But as we know saturation momentun $Q_s$ is not a constant. It depends on Bjorken variable $x$ which in this process defined by relation
\begin{equation}
	x=\frac{q}{\sqrt{s}}
\end{equation}
and as the only known scale is $q_C$ then instead of (\ref{pp_simple}) we'll have 
\begin{equation}
	q_C=\beta Q_s\left(\frac{\beta_{1}q_C}{\sqrt{s}}\right)
	\label{p_cc}
\end{equation}
where $\beta_{1}$ is another dimentionless constant.

It is easy to define $Q_s(x)$. From geometric scaling effect for small $x$ we have
\begin{equation}
Q_s^2(x)=A^{1/3}Q_{s0}^2{\left(\frac{x_0}{x}\right)}^{\lambda}
\end{equation}
where $\lambda=0.3$ is geometric scaling constant and $Q_{s0}, x_0$ some parameters those exact value we define from
fact that for reaction with $Au$ nucles at $\sqrt{s}=200 \, Gev$ we have $Q_s=1-2 \, Gev$.
It is easy to solve (\ref{p_cc}) and write expression for $q_C$
\begin{equation}
	q_C=q^0_{C}A^{\frac{1}{3(2+\lambda)}}{\sqrt{s}}^{\frac{\lambda}{2+\lambda}}
\end{equation}
or if we log both parts we'll have:
\begin{equation}
	\ln(q_C)=a+b\ln(\sqrt{s})
	\label{ex_qc}
\end{equation}
where $a$ and $b$ defined as:
\begin{eqnarray}
	\label{exact_q_c}
	a=\frac{1}{3(2+\lambda)}ln(A)+ln(q^0_C) \\\nonumber
	b=\frac{\lambda}{2+\lambda}=0.1304
\end{eqnarray}
So to test saturation model prediction we should calculate Cronin momentum for different energies and check
(\ref{ex_qc}). There is however soft scale $\Lambda_{QCD}$ and it is not obvious that it's existance does not change this formula.
So we should check this dependence explicitly by numerical calculation.
Let us consider Cronin ratio for $p+A$ (from here we take $Au$ nucleus) collision
\begin{equation}
\label{dddddd}
    R_{pA}=\frac{\frac{d\sigma^{pA}}{dyd^2q}}{A\frac{d\sigma^{pp}}{dyd^2q}}
\end{equation}
As we stated before we suppose $y=0$.
In saturation model gluon production cross-section can be expressed as
\begin{equation}
\frac{d \sigma^{pA}}{d^2 q \ dy} \, = \, \frac{2 \, \as}{C_F} \,  
\frac{1}{q^2} \, \int d^2 k \, \phi_p (x_1,q^2) \, \phi_A (x_2,(q  
- k)^2),  
\label{section_exact}
\end{equation}
where $\phi_{A,p}$ is unintegrated gluon distribution of nucleus and proton and
$x_1,x_2$ defined by equation
\begin{eqnarray}
x_1=\frac{q}{\sqrt{s}}e^{-y},\ x_2=\frac{q}{\sqrt{s}}e^{y}
\end{eqnarray}
In leading logarifmic order we can rewrite (\ref{section_exact}) in following form \cite{gribov}:
\begin{equation}
\frac{d\sigma^{pA}}{d^2 q \ dy} \, = \, \frac{2 \, \as}{C_F} \,  
\frac{1}{q^2} \, \left(
x_1G_p(x_1,q^2)\phi_A(x_2,q^2)+
x_2G_A(x_2,q^2)\phi_p(x_1,q^2)
\right),
\end{equation}
where $xG(x,q^2)$ is gluon distributin function which can be expressed from $\phi(x,k^2)$ by following relation
\begin{equation}
    xG(x,q^2)=\int_{\Lambda}^q{ \phi(x,k^2)dk^2}
\end{equation}
Using the same approximation we can write following approximate relation for cross-section of gluon production in $p+p$ collisions
\begin{equation}
\frac{d\sigma^{pp}}{d^2 a \ dy} \, = \, \frac{2 \, \as}{C_F} \,  
\frac{1}{k^2} \, \left(
x_1G_p(x_1,a^2)\phi_p(x_2,q^2)+
x_2G_p(x_2,q^2)\phi_p(x_1,q^2)
\right),
\end{equation}
Let us suppose that in considered kinematical region unintegrated gluon 
distibution function of proton $\phi_p(x,q^2)$ does not depend on $x$ and that
$\phi_p(x,q^2)=\frac{\as C_F}{\pi}\frac{1}{q^2}$.
Then Cronin ratio can be expressed as:
\begin{equation}
R_{pA}=\frac{1}{A}
\left(
\frac{\phi_A(x_2,q^2)}{\phi_p(x_2,q^2)}+
\frac{G_A(x_2,q^2)}{G_p(x_2,q^2)}
\right)
\end{equation}
or as we supposed $y=0$ then
\begin{equation}
	R_{pA}=\frac{1}{A}
	\left(
	\frac{\phi_A(x,q^2)}{\phi_p(x,q^2)}+
	\frac{G_A(x,q^2)}{G_p(x,q^2)}
	\right)
\label{cronin_ratio_app}
\end{equation} where $x=\frac{q}{\sqrt{s}}$

All we need now is the expression for unintegrated gluon distribution function. We consider three models for gluon distribution function:
Kharzeev-Levin-Nardi proposed in \cite{GDF_KL},
MacLerran-Venugopalan proposed in \cite{GDF_ML,GDF_ML2} and "dipole" model.

\section{Kharzeev-Levin-Nardi model}
In simplified form of this model unintegrated gluon distribution function $\phi(x,q^2)$ have following form:
\begin{eqnarray}
    \phi_A(x,q^2)=\phi_A^0, q<Q_s(x)\\\nonumber
    \phi_A(x,q^2)=\phi_A^0\frac{Q_s^2(x)}{q^2}, q>Q_s(x)\nonumber,
\end{eqnarray}
where $\phi_A^0$ normalization factor.

\begin{figure}
\hbox{ \epsffile{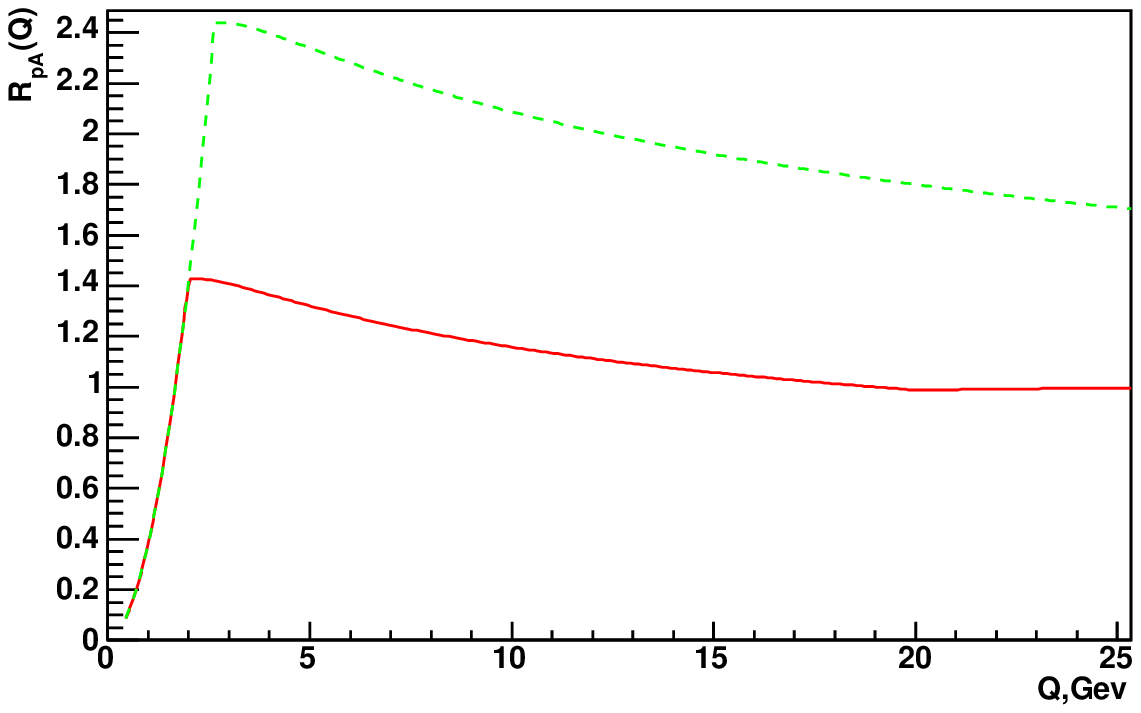}}
\caption{Cronin ratio for Kharzeev-Levin-Nardi gluon distribution function
for $p+A$ collisions $\sqrt{s}=200 Gev$(solid curve) and 
$\sqrt{s}=1700 Gev$(dashed curve)}
\label{fig_KL_pa}
\end{figure}

Then for gluon distribution function $G(x,q^2)$ we have:
\begin{eqnarray}
    xG_A(x,q^2)=\phi_A^0\left(q^2-\Lambda^2_{QCD}\right), q<Q_s(x)\\\nonumber
    xG_A(x,q^2)=\phi_A^0\left(Q^2_s(x)ln\left(\frac{q^2}{Q^2_s(x)}\right)+
    Q^2_s(x)-\Lambda^2_{QCD}\right), q>Q_s(x)
\end{eqnarray}
And for Cronin ratio we have(we use here approximate formula (\ref{cronin_ratio_app}))
\begin{eqnarray}
R_{pA}=\frac{1}{A}
\phi_A^0\frac{\pi}{\as C_F}
\left(q^2+\frac{q^2-\Lambda^2_{QCD}}
{ln\left(\frac{q^2}{\Lambda^2_{QCD}}\right)}\right), q<Q_s(x)\\
R_{pA}=\frac{1}{A}
\phi_A^0\frac{\pi}{\as C_F}
\left(Q^2_s(x)+\frac{Q_s^2(x)ln\left(\frac{q^2}{Q^2_s(x)}\right)+
    \left(Q^2_s(x)-\Lambda^2_{QCD}\right)}
    {ln\left(\frac{q^2}{\Lambda^2_{QCD}}\right)}
\right), q>Q_s(x)
\label{Cronin_KL}
\end{eqnarray}
If we look at (\ref{Cronin_KL}) we'll see that $R_{pA}$ here is non-decreasing function of momentum $q$ so is not clean
if there is any Cronin like behavior in this model. It should be mentioned however that $x$ depends on $q$ by means of relation 
$x=\frac{q}{\sqrt{s}}$ (if we suppose $y=0$) and as saturation momentum in low $x$ region depend on $x$ as
$Q^2_s(x)=A^{1/3}Q^2_{s0}\left(\frac{x_0}{x}\right)^\lambda$ then in reality (\ref{Cronin_KL}) have maximun at some 
monentum $q_C$ (figure \ref{fig_KL_pa}) which value {\it approximately} defined by equation
\begin{equation}
q_C=Q_s\left(\frac{q_C}{\sqrt{s}}\right)
\end{equation}
and modified slightly by logarifmic terms in (\ref{Cronin_KL}). 
Nevertheless we have formula similair to (\ref{ex_qc}):
\begin{equation}
   \ln(q_C)=a+b\ln(\sqrt{s})
\end{equation}
where $b=0.1042$
\section{McLerran-Venugopalan model}
In McLerran-Venugopalan model expression for unintegrated gluon distribution function was finded in works 
\cite{GDF_ML,GDF_ML2} and can be written as
\begin{equation}
\phi_A (x,  q^2) \, =  \,
\frac{4 \, C_F}{\as \, (2 \pi)^3} \, \int d^2 b \,
d^2 r \, e^{- i q \cdot { r}} \ \frac{1}{{ r}^2} \ (1 -
e^{-{ r}^2 Q_{s}^2 \ln(1/r\Lambda) /4}),
\end{equation}

Or if consider cilindrical nucleus

\begin{equation}
\phi_A (x, q^2) \, =  \,
\frac{4 S_A\, C_F}{\as \, (2 \pi)^3} \int \,
d^2 r \, e^{- i q \cdot { r}} \ \frac{1}{{ r}^2} \ (1 -
e^{-{ r}^2 Q_{s}^2 \ln(1/r\Lambda) /4}),
\end{equation}
or 

\begin{equation}
\phi_A (x, q^2) \, =  \,
\frac{4 S_A\, C_F}{\as \, (2 \pi)^2} \int \,
dr \, J_0(q \cdot r) \ \frac{1}{r} \ (1 -
e^{- r^2 Q_{s}^2 \ln(1/r\Lambda) /4}),
\end{equation}
It is better however use the expression proposed in \cite{Kharz_111} which relates unintegrated gluon distribution
function in McLerran-Venugopalan model and the forward amplitude of scatering $N_G(r, x)$ of a gluon dipole of transverse
size $r$ and rapidity $y=ln(1/x)$ on nucleus. When we can rewrite previous equation in the following form:
\begin{equation}
\phi_A (x, q^2) \, =  \,
\frac{4 S_A\, C_F}{\as \, (2 \pi)^2} \int \,
dr \, J_0(q \cdot r) \ \frac{1}{r} \ N_G(r,x),
\end{equation}
where $J_0(x)$ is Bessel function.
It obvious that $N_G$ depends on $x$. There are different ways to set this dependence.
We can use Balitsky-Kovchegov equation to define $x$ dependence of dipole scattering cross-section but as it is unsolved for now
we choise more simple way.
Let us define {\it ad hoc} that
\begin{equation}
N_G(r,x)=1-e^{-r^2Q_s(x)^2ln(1/r\Lambda)/4}
\label{n_g}
\end{equation}
i.e. all $x$ dependence goes in definition of $Q_s(x)$.
But we can not use (\ref{n_g}) directly as $N_G(r,x)$ have not very good behavior for large $r$( i.e. if $r\rightarrow\infty$ then $N_G(r,x)$ becomes negative instead of unity). So we should regularize (\ref{n_g}) somehow.
Let us regularize $N_G(r,x)$ by following prescription:
\begin{equation}
N_G(r,x)=1-e^{r^2Q_s(x)^2(ln(r\Lambda)-\sqrt{(\ln(r\Lambda))^2+\epsilon^2)}+\ln(r_0\Lambda))/8}
\end{equation}
and set $r_0=\frac{1}{\sqrt{e}\Lambda}$ and $\epsilon<1$(the final result does not depend on exact value of $\epsilon$, if it is not too large).
It should be noted that result does not depend on regularization scheme and
we could regularize $N_G(r,x)$ with something like this:
\begin{eqnarray}
N_G(r,x)=1-e^{-r^2Q_s(x)^2ln(1/r\Lambda)/4} \  , r<r_0 \\\nonumber
N_G(r,x)=1-e^{-r^2Q_s(x)^2ln(1/r_0\Lambda)/4} \ , r>r_0
\end{eqnarray}
but this regularization is inconvenient in "dipole" model.

\begin{figure}
\hbox{\epsffile{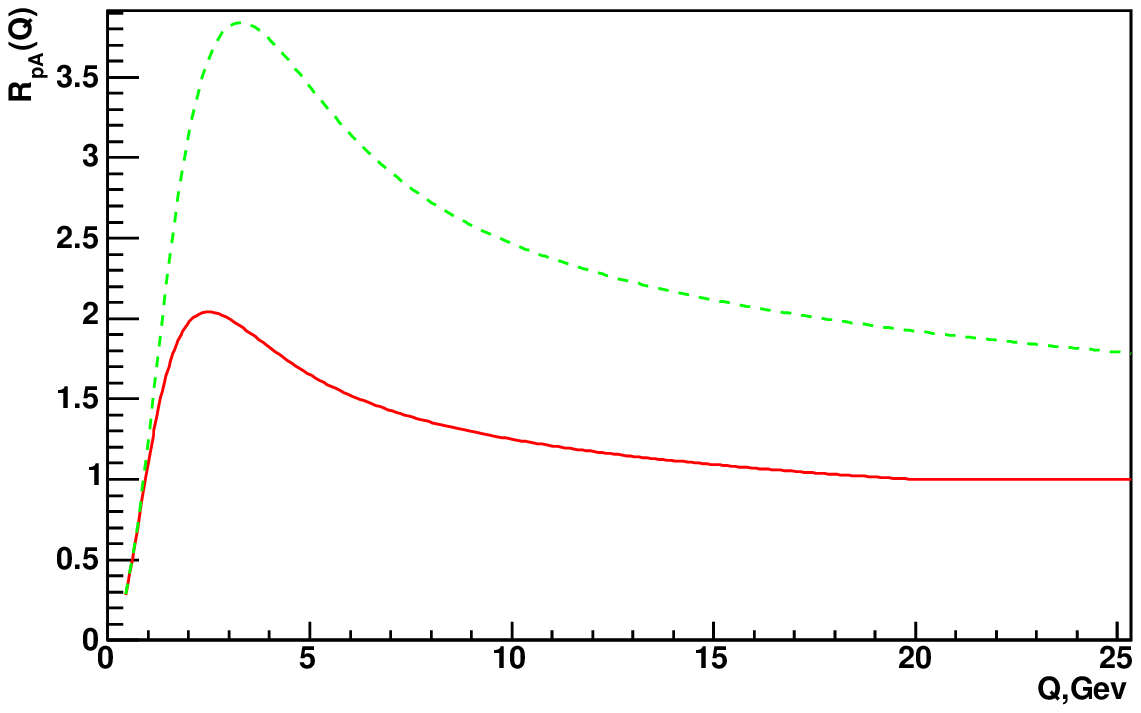}}
\caption{Cronin ratio for McLerran-Venucopalan gluon distribution function
for $p+A$ collisions for $\sqrt{s}=200 Gev$(solid curve) and 
$\sqrt{s}=1700 Gev$ (dashed curve)}
\label{fig_ML_pa}
\end{figure}

\begin{figure}
\epsffile{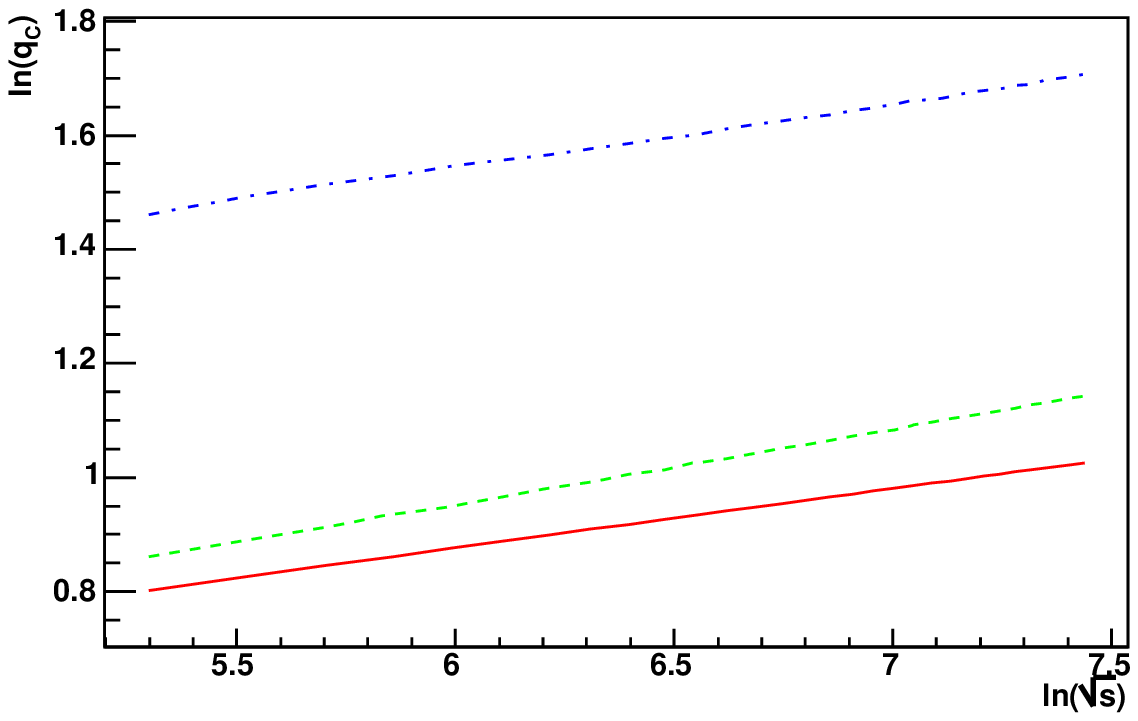}
\caption{Dependence of $ln(q_C)$(for $R_{pA}$)on $ln(\sqrt{s})$ for different models:
Kharzeev-Levin-Nardi(solid curve), McLerran-Venugopalan(dashed curve), "dipole"
(dot-dashed curve)}
\label{fig_q_s_pa}
\end{figure}

Then for gluon distribution function $G(x,q^2)$ we have:
\begin{equation}
xG_A (x, q^2) \, =  \,
\frac{4 S_A\, C_F}{\as \, (2 \pi)^3} 2 \int \,
dr \, (qJ_1(q \cdot r)-\Lambda J_1(\Lambda \cdot r)) \ \frac{1}{r^2} \ N_G(r,x)),
\end{equation}
If we substitute this functions in Cronin ratio (\ref{cronin_ratio_app}) we'll have dependence which is presented 
in figure \ref{fig_ML_pa}.
Then we can calculate numericaly value of Cronin momentum $q_C$ for different energies.
The result is presented in figure \ref{fig_q_s_pa}.
The slope is $b=0.1323$.
It should be mentioned that even the line have different position they have almost the same slope as previous model
and almost exactly equal one which was calculated in (\ref{exact_q_c}).
\section{"Dipole" model}
In "dipole" model we can relate unintegrated gluon distribution function with gluon dipole cross-section.
It was done in work \cite{Braun1,Braun2} and expression for unintegrated gluon distribution function can be written as (we supposed that nucleus is cilindrical)
\begin{equation}
\phi_A (x, q^2) \, =  \,
\frac{4 S_A\, C_F}{\as \, (2 \pi)^3} \int \,
d^2 r \, e^{- i q \cdot { r}} \ \nabla^2_r \ N_G(r,x)),
\end{equation}
or
\begin{equation}
\phi_A (x, q^2) \, =  \,
\frac{4 S_A\, C_F}{\as \, (2 \pi)^2} \int \,
dr \, J_0(q \cdot r) \ r\nabla^2_r \ N_G(r,x),
\end{equation}

\begin{figure}
\epsffile{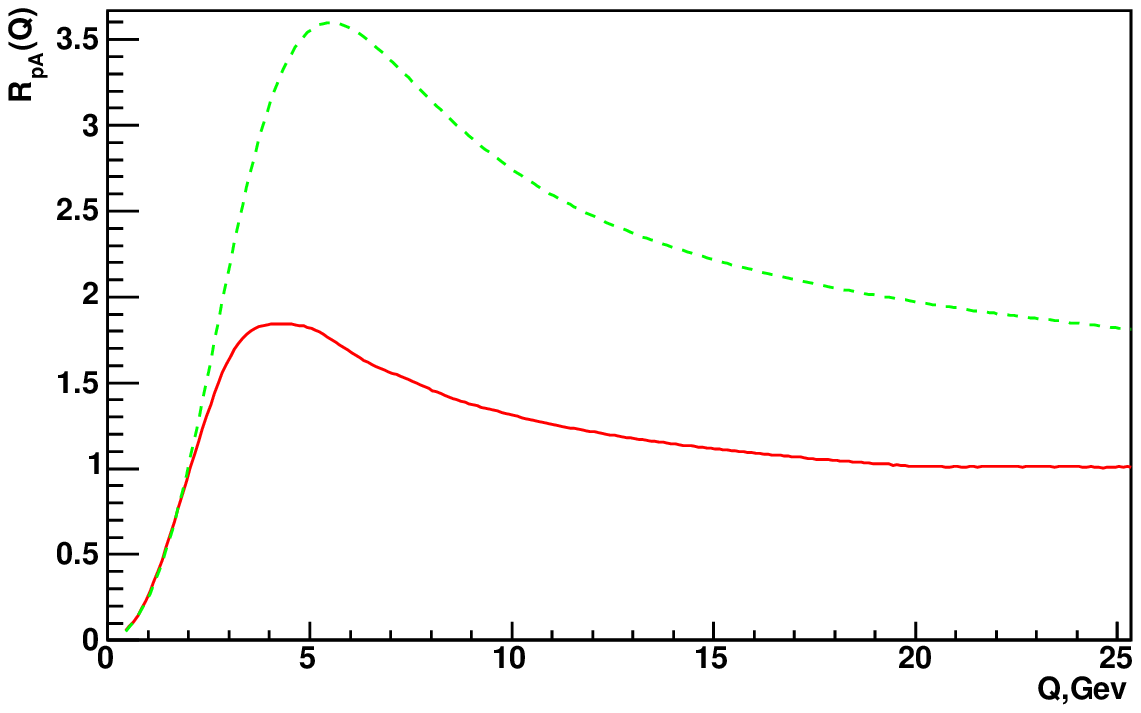}
\caption{Cronin ratio for "dipole" gluon distribution function
for $p+A$ collisions for $\sqrt{s}=200 Gev$(soild curve) and 
$\sqrt{s}=1700 Gev$ (dashed curve)}
\label{fig_dip_pa}
\end{figure}

For for gluon distribution function $G(x,q^2)$ we'll have
\begin{equation}
xG_A (x, q^2) \, =  \,
\frac{4 S_A\, C_F}{\as \, (2 \pi)^2} 2 \int \,
dr \, (qJ_1(q \cdot r)-\Lambda J_1(\Lambda \cdot r)) \ \nabla^2_r N_G(r,x),
\end{equation}
As before it is easy to calculate numerically value of Cronin ratio. Result is presented on figure \ref{fig_dip_pa}.
Varying energy we calculate numerically Cronin momentum $q_C$ (result presented on figure \ref{fig_q_s_pa}).
And as before we have dependence $\ln(q_C)=a+b\ln(\sqrt{s})$, with slope $b=0.1120$
\section{A+A collisions}
\begin{figure}
\hbox{ \epsffile{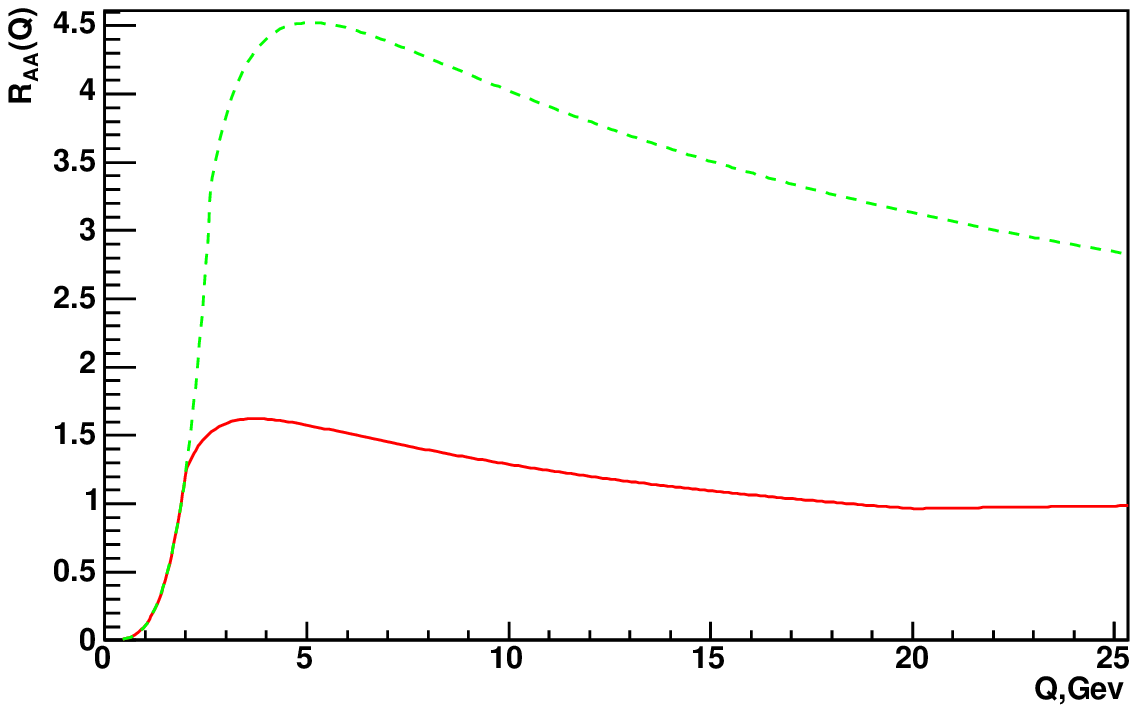}}
\caption{Cronin ratio for Kharzeev-Levin-Nardi gluon distribution function
for $A+A$ collisions $\sqrt{s}=200 Gev$(solid curve) and 
$\sqrt{s}=1700 Gev$(dashed curve)}
\label{fig_KL_aa}
\end{figure}

\begin{figure}
\epsffile{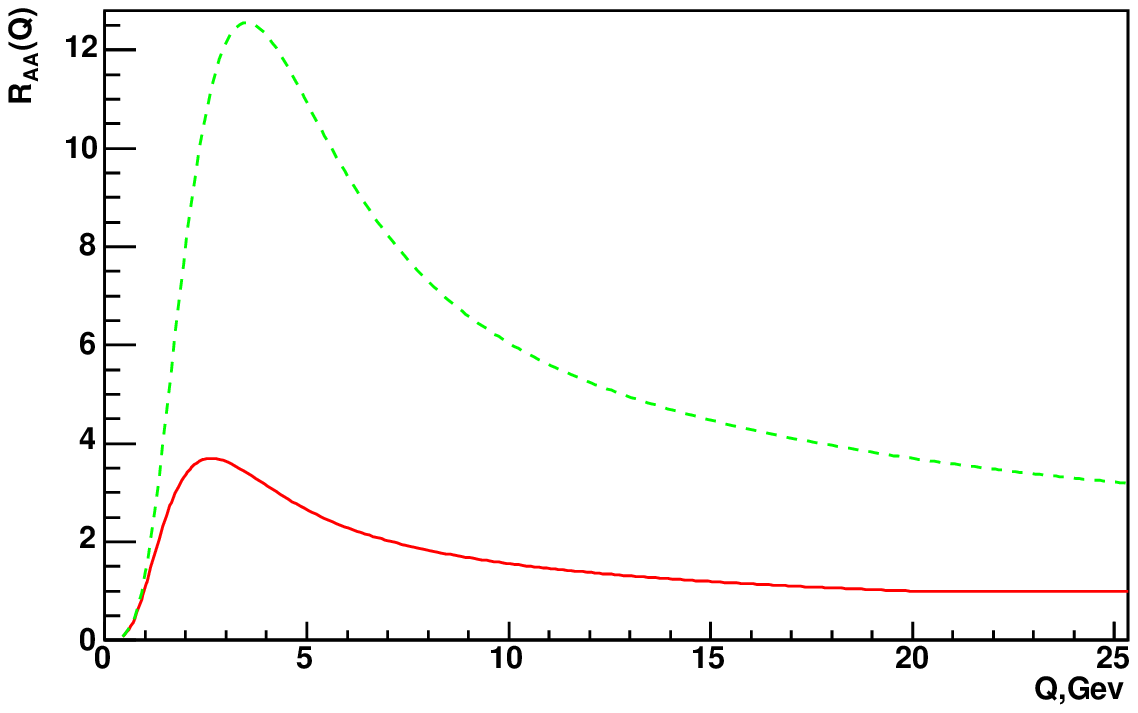}
\caption{Cronin ratio for McLerran-Venucopalan gluon distribution function
for $A+A$ collisions for $\sqrt{s}=200 Gev$(solid curve) and 
$\sqrt{s}=1700 Gev$ (dashed curve)}
\label{fig_ML_aa}
\end{figure}

\begin{figure}
\epsffile{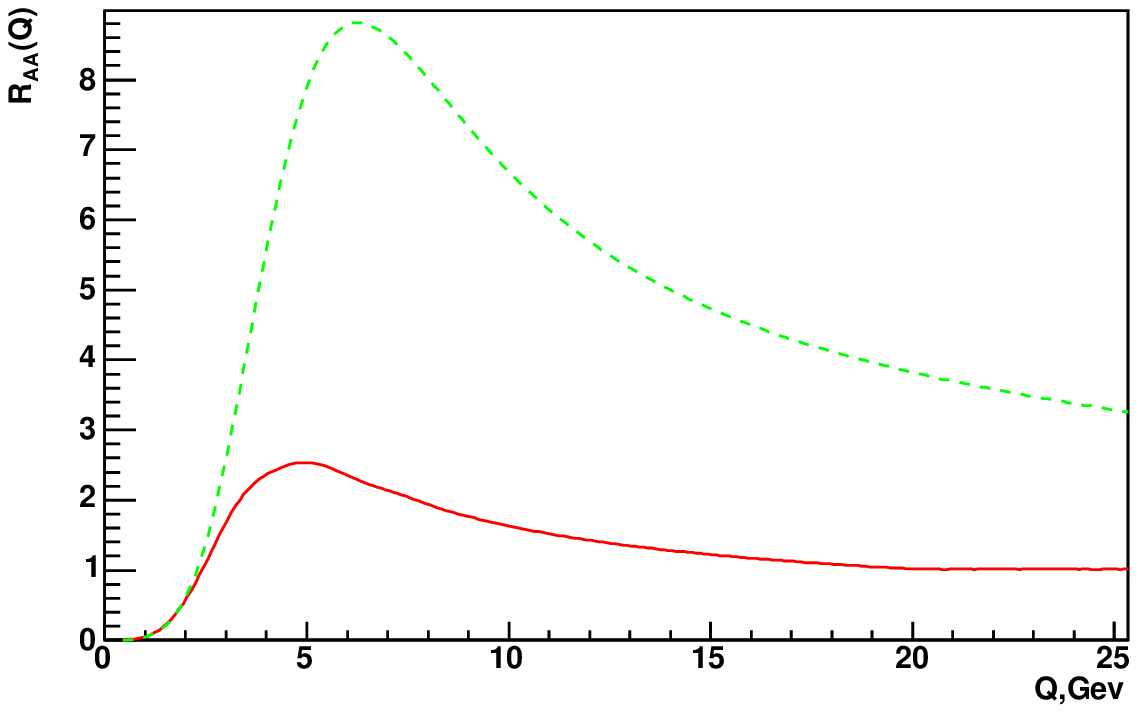}
\caption{Cronin ratio for "dipole" gluon distribution function
for $A+A$ collisions for $\sqrt{s}=200 Gev$(solid curve) and 
$\sqrt{s}=1700 Gev$ (dashed curve)}
\label{fig_dip_aa}
\end{figure}

\begin{figure}
\epsffile{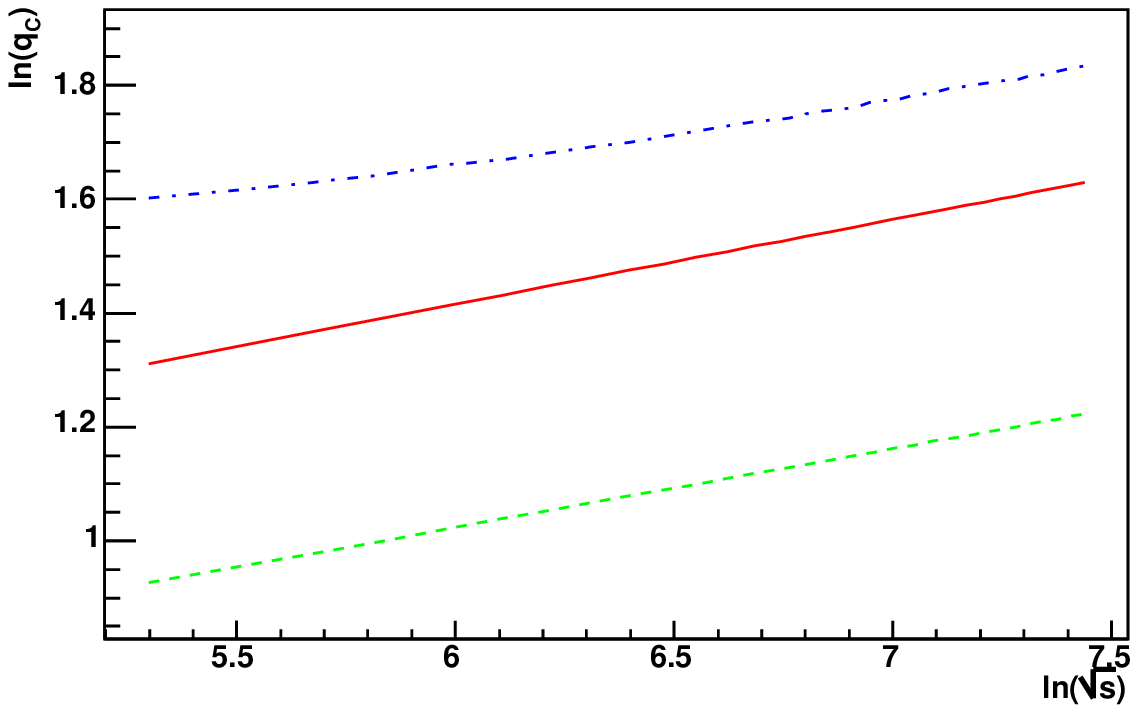}
\caption{Dependence of $ln(q_C)$(for $R_{AA}$) on $ln(\sqrt{s})$ for different models:
Kharzeev-Levin-Nardi(solid curve), McLerran-Venugopalan(dashed curve), "dipole"
(dot-dashed curve)}
\label{fig_q_s_aa}
\end{figure}

Like in $p+A$ collision in $A+A$ collisions(we take only central rapidity region) there is only one semihard scale $Q_s$.
Ant therefore dependence of Cronin momentum $q_C$ must be govered by (\ref{p_cc}).
We can apply all formulas above to this case, as we have for Cronin ratio following approximate relation similair to (\ref{cronin_ratio_app})
\begin{equation}
	R_AA=\frac{G_A(x,p)}{G_p(x.p)}\frac{\phi_A(x,p)}{\phi_p(x,p)}
\end{equation}
calculating numerically $q$ dependence of Cronin ration for considered models 
(figures \ref{fig_KL_aa},\ref{fig_ML_aa},\ref{fig_dip_aa}) at differen energies we have same linear behavior for $ln(q_C)$ as
before (figure \ref{fig_q_s_aa}) and also have slopes consistent with (\ref{exact_q_c}). All data summarized in Table~1 (for 'dipole' model only points with $\sqrt{s}>500 Gev$ was taked for slope calculation).

\begin{table}
\center
\label{tab:mod}
\begin{tabular}{|l|c|c|}\hline
                 & {$p+A$} &  {$A+A$} \\
Model           &  & \\ \hline
Kharzeev-Levin-Nardi & 0.1042 & 0.1485 \\ 
McLerran-Venugopalan & 0.1323 & 0.1383 \\ 
'Dipole'             & 0.1120 & 0.1244 \\ \hline
\end{tabular}
\caption{Summary of slopes for different models in $p+A$ and $A+A$ collisions.}
\end{table}

\section{Conclusion}
We calculate $\sqrt{s}$ dependence of Cronin momentum in several models based on saturation and show that this 
dependence is consistent with simple formula based on geometric scaling only. This subtle prediction can test validity of
saturation model. Even more. As slope values is slightly different we have posibility to distinguish among variants.
But this requires more precice measurement of Cronin effect(at least in midle momentum region) that those we have today.
Having this we can in turn measure dependence of saturation momentum $Q_s(x)$ on $x$.

\section*{Acknowledgments}
We thank A. Dmitriev and A. Popov for useful discussions.
This work was supported by RFBR Grant RFBR-03-02-16157a and grant of Ministry for Education E02-3.1-282

\newpage

\end{document}